\documentclass{PoSb}

\usepackage{epsfig,graphics,amsmath,bbm}
\newcommand{\rhs}{right-hand side}
\newcommand{\naive}{na\"{\i}ve}

\newcommand{\hepth}[1]{hep-th/#1}
\newcommand{\hepph}[1]{hep-ph/#1}
\newcommand{\condmat}[1]{cond-mat/#1}

\newcommand{\eq}[1]{(\ref{eq:#1})}

\newcommand{\fig}[1]{figure~\ref{fig:#1}}

\newcommand{\be}{\begin{equation}}
\newcommand{\ee}{\end{equation}}

\newcommand{\bcf}{\begin{center}\begin{figure}}
\newcommand{\ecf}{\end{figure}\end{center}}

\newcommand{\cdeps}[1]{\ensuremath{\begin{array}{c}\includegraphics{#1.eps} \end{array}}} 

\newcommand{\flow}{\Lambda \partial_\Lambda}
\newcommand{\cutoff}{K}
\newcommand{\ep}{C}
\newcommand{\dd}{\dot{C}}

\newcommand{\Stot}{S}
\newcommand{\Sint}{S^{\mathrm{I}}}
\newcommand{\hS}{\hat{S}}

\newcommand{\knl}[1]{\cdot {#1}\cdot}
\newcommand{\classical}[4]{\fder{#1}{#4} \knl{#2} \fder{#3}{#4}} 
\newcommand{\quantum}[3]{\fder{}{#3} \knl{#1} \fder{#2}{#3§}}

\newcommand{\pder}[2]{\frac{\partial #1}{\partial #2}}
\newcommand{\fder}[2]{\frac{\delta #1}{\delta #2}}

\newcommand{\D}{d}
\newcommand{\volume}[1]{d^{\D} \! #1 \,}
\newcommand{\Int}[1]{\int \!\! \volume{#1}}
\newcommand{\Fint}[1]{\int \mathcal{D} #1 \,}
\newcommand{\MomInt}[2]{\int \!\! \frac{d^{#1} #2}{(2\pi)^{#1}} \, }

\newcommand{\hf}{\frac{1}{2}}

\newcommand{\Or}{\mathrm{O}}
\newcommand{\order}[1]{\Or \bigl( #1 \bigr)}

\newcommand{\SU}{\mathrm{SU}}
\newcommand{\tr}{\mathrm{tr}\,}
\newcommand{\str}{\mathrm{str}\,}

\title{Aspects of Manifest Gauge Invariance}

\ShortTitle{Aspects of Manifest Gauge Invariance}

\author{\speaker{Oliver J.~Rosten}\\
        Department of Physics and Astronomy, University of Sussex, Brighton, BN1 9QH, U.K.\\
        E-mail: \email{O.J.Rosten@Sussex.ac.uk}}

\abstract{The essential elements of manifestly gauge invariant exact renormalization groups are recalled. The pros and cons of the formalism are discussed and it is argued that now is the right time to try to utilize the formalism to provide new insights into the strong coupling domain of QCD.}

\FullConference{The many faces of QCD\\
		November 2-5, 2010\\
		Gent Belgium}

\begin{document}

\section{Introduction}

\subsection{Overview}

For some time now, there has existed a local, manifestly gauge invariant approach to QCD, formulated directly in the continuum~\cite{aprop,mgierg1,qcd}. Whilst the scheme has been well tested perturbatively, its apparent complexity has impeded its widespread use. However, recent developments in the understanding of the underlying formalism suggest that, in the near future, it will be profitable to revisit this method, with the particular aim of realizing the original plan to apply it in the strong-coupling domain.

The approach is based upon the Exact Renormalization Group (ERG) which, since its christening by
Wilson \& Kogut~\cite{Wilson}, has provided a robust approach to dealing with a wide range of intrinsically nonperturbative problems~\cite{Wetterich-Rev,B+B,JMP-Review,Gies-Rev,Delamotte-Rev,Fundamentals}. It is partly because of this success that it is hoped that manifestly gauge invariant ERGs will have something to offer in the challenging arena of the low energy physics of Yang-Mills. 

Indeed, there has already been quite some success in applying the ERG to both pure Yang-Mills and QCD. However, this has been achieved within an implementation where the cutoff fundamental to the approach breaks gauge invariance. Formally, this breaking disappears in the limit that all quantum fluctuations are integrated out, though this property is spoiled by truncations necessary to perform actual calculations. Nevertheless, the formalism is comparatively easy to use and, moreover, one can hope to keep the errors induced by truncations under control. See~\cite{Gies-Rev} for a balanced discussion of the pros and cons of the approach. From a practical perspective, the formalism has given encouraging and impressive results, 
with some of the more recent ones described by Pawlowski in~\cite{JMP-PhaseDiagram-Conf} and these proceeding.

However, from a conceptual point of view, the advantages of an ERG equation for which gauge invariance is preserved all the way along the flow are clear; that gauge invariance can even be left manifest is an added bonus. The drawback, as mentioned above, is the complexity of the approach. The game, then, is to attempt to simplify things to such a level where the benefits of manifest gauge invariance win. Already, it has been understood that manifestly gauge invariant ERGs possess a hidden simplicity~\cite{mgiuc,univ}. Now the task is to develop this further and, in particular, to understand how to exploit it for nonperturbative calculations.

\subsection{Why the Exact Renormalization Group?}

The ERG grew out of the pioneering work of Wilson~\cite{Wilson}, and others, into the study of systems exhibiting a large number of degrees of freedom per correlation length. Wilson realized that, so long as the fundamental interactions are suitably local, then an understanding of such systems can be built up by an iterated application of what is essentially Kadanoff's blocking procedure~\cite{Kadanoff}. The logic is as follows. First, start by partitioning up the system of interest into small subsystems, ideally containing just a few degrees of freedom. A selection of such subsystems is shown in the first panel of \fig{iterate}. If the correlation length were small (i.e.\ the same as the characteristic size of one of these subsystems), then we would essentially be done since the properties of the entire system would be basically the same as those of a subsystem. However, we suppose that we are not in this situation. To proceed, let us restrict ourselves to the case where the interactions are local; in this case 
the subsystems essentially talk only to their neighbours. With this in mind, group the subsystems into blocks, as shown in the second panel of \fig{iterate}. Now, since it is supposed that each individual subsystem is easy to understand, in the sense that it possesses only a small number of degrees of freedom, and that subsystems only talk to their neighbours, we can hope `coarse-grain' over blocks, to give an effective description in terms of bigger subsystems, as indicated in the final panel of \fig{iterate}. With this done, we have an understanding of the physics at the scale of the coarse-grained subsystems, which is closer to the correlation length than where we started. Moreover, once again, we have a description in terms of subsystems which are comparatively easy to understand and which effectively only talk to their neighbours. The key point is that this procedure can be iterated,  allowing us to systematically build up an understanding of the system at scales of order the correlation length. 
\bcf[h]
	\[
		\resizebox{10cm}{!}{\cdeps{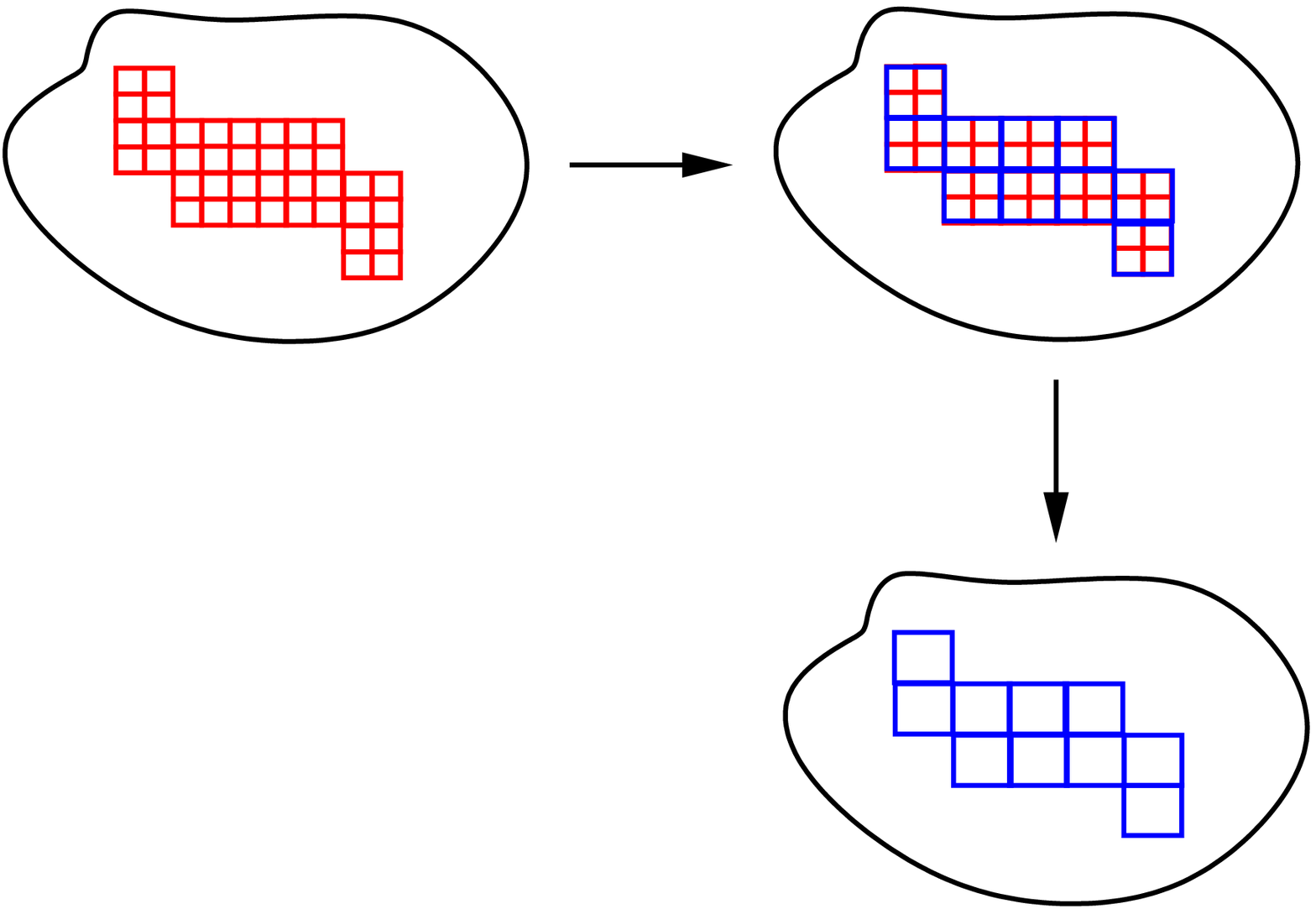}}
	\]
\caption{Coarse-graining over subsystems in order to build up an understanding of systems with a large number of degrees of freedom per correlation length.
In the first panel, the system is divided up into subsystems. In the second panel, neighbouring subsystems are grouped together into `blocks'. Next, blocks are coarse-grained to give a description of the system in terms of effective degrees of freedom.
}
\label{fig:iterate}
\ecf

With this compelling picture in our minds, let us emphasise that the coarse-graining is not done at the level of physical samples. Rather, the idea is to coarse-grain over degrees of freedom within the partition function, thereby obtaining a succession of actions which provide effective descriptions of the physics at the appropriate scale.

\subsection{Applying the Exact Renormalization Group to Quantum Field Theory}

We have learnt that the ERG is of use for describing systems exhibiting both a large number of degrees of freedom per correlation length and locality. Consequently, it is reasonable to apply the formalism in the context of quantum field theory, which we formulate in $\D$ Euclidean dimensions. 

The starting point is the partition function, which we suppose to possess some overall cutoff, $\Lambda_0$ (the bare scale). We now consider integrating out degrees of freedom down to some effective scale, $\Lambda$, such that the partition function remains the same. During this process, the bare action evolves into the Wilsonian effective action, $\Stot_\Lambda$. Of course, for this to work, the Wilsonian effective action must satisfy a consistency condition, which is encoded in an ERG equation. This equation tells us how $\Stot_\Lambda$ changes under infinitesimal changes of the scale, viz.\
\[
	-\flow \Stot_\Lambda = \ldots
\]
The form of the \rhs\ will depend on the precise way in which we coarse-grain over degrees of freedom. The presence of this flexibility, which will be discussed further below, should be clear from \fig{iterate}:
the choice to group subsystems into $2\times2$ blocks was an arbitrary one, and this arbitrariness is one manifestation of the freedom to choose different coarse-graining schemes.

Before giving technical details, let us note that we can immediately detect a potential difficulty in applying this approach to gauge theories. A central element of the ERG is a cutoff, which partitions modes into those of high and low energies. However, for gauge theories, one needs to be careful since a \naive\ partitioning is not gauge invariant (except in the largely uninteresting case of pure Abelian gauge theory). To rectify this problem amounts to finding a regularization of gauge theory based on a cutoff. Fortunately, such a scheme---which combines covariant higher derivative and Pauli-Villars regularizations---has been constructed~\cite{sunn,qcd}, as we will recall below.

\section{Exact Renormalization Group Equations}

\subsection{General Considerations}

There are several different routes to obtaining ERG equations. In order to be sympathetic to the discussion above, we will highlight the role played by blocking. Starting at the bare scale, let us denote our bare field by $\varphi_0$ (we need not suppose that this is a scalar field). We now introduce the `blocking functional', $b_\Lambda$, which serves to coarse-grain degrees of freedom over patches with a characteristic size $1/\Lambda$. Thus, the blocked field, $\varphi$, is written as
\be
	\varphi(x) = b_\Lambda[\varphi_0](x).
\label{eq:block-field}
\ee
A sensible choice would be $b_\Lambda[\varphi_0](x) = \Lambda^d \Int{y} f((x-y)\Lambda)\varphi_0(y)$, where $f(z\Lambda)$ decays rapidly for $z\Lambda >1$. However, there are many other valid choices for $b$: indeed, there is no need for the blocking to be linear in the field. 

Of course, the fact that blocking is performed over patches of finite size introduces a degree of non-locality. However, there are strong restrictions on the type of non-locality allowed. In particular, in the limit $\Lambda \rightarrow \infty$, everything must become strictly local. More generally, in momentum space, $f(p^2/\Lambda^2)$ is, for small argument, required to be analytic. Such behaviour is referred to as quasi-locality and is a crucial property of the Wilsonian effective action: generally speaking, we are interested in solving the ERG equation subject to the condition that the solutions are quasi-local.

By insisting on quasi-locality, we ensure that, for theories which are renormalizable in the Wilsonian sense, the Wilsonian effective action encodes the physics of a \emph{local} QFT. This can be understood as follows. Renormalizability implies that we can take the limit $\Lambda \rightarrow \infty$. But in this limit, quasi-locality ensures that the action becomes strictly local. Now, recall that things are set up such that the partition function is invariant under the flow. Therefore, the Wilsonian effective action at the scale $\Lambda$ describes precisely the same partition function as the local action in the deep ultraviolet (UV) limit, where the flow is spawned.

To complete the picture, note that `genuine' non-locality (in the sense of non-analyticity) is permitted to arise in the limit $\Lambda \rightarrow 0$. This makes sense. In this limit, we have integrated out all quantum fluctuations and, moreover, expect the partition function to contain information about the correlation functions (which is, of course, most readily extracted in the presence of a source). Of course, it is quite permissible for correlation functions to be non-local. Note, though, that for any $\Lambda >0$, we have not yet integrated out all quantum fluctuations. This is equivalent to saying that the modes below $\Lambda$ have been suppressed, which can be understood in terms of an infrared (IR) cutoff. Only in the limit that this cutoff is removed do we expect non-locality (as opposed to quasi-locality) to emerge. Below, we will give an explicit example of a solution to the flow equation exhibiting this transition from strict locality to quasi-locality to non-locality.

The aim now is to go from the blocking functional to an explicit ERG equation. First of all, let us note from~\eq{block-field} that we can relate the Wilsonian effective action to the bare action as follows:
\be
	e^{-\Stot_\Lambda[\varphi]} = \Fint{\varphi_0} \delta\bigl[ \varphi - b_\Lambda[\varphi_0]\bigr]
	e^{-\Stot_{\Lambda_0}[\varphi_0]}.
\label{eq:blocking}
\ee
Now, rather than specifying $b_\Lambda$ explicitly, it turns out to be more convenient to do so implicitly, via
\be
	\Psi(x) e^{-\Stot_\Lambda[\varphi]}
	=
	 \Fint{\varphi_0} \delta\bigl[ \varphi - b_\Lambda[\varphi_0]\bigr]
	 \Lambda \pder{b_\Lambda[\varphi_0](x)}{\Lambda}
	e^{-\Stot_{\Lambda_0}[\varphi_0]},
\ee
which serves as a definition of $\Psi$. From this, it follows that
\be
	-\flow e^{-\Stot_\Lambda[\varphi]} =  \Int{x} \fder{}{\varphi(x)} 
	\Bigl\{
	\Psi(x) e^{-\Stot_\Lambda[\varphi]}
	\Bigr\},
\label{eq:blocked}
\ee
where the derivative with respect to $\Lambda$ is performed at constant $\varphi$.
Thus, the freedom to choose the blocking functional, $b_\Lambda$, has been translated into a choice of $\Psi$. It is worth noting that this equation can alternatively be derived by supposing that, under an RG step $\Lambda \rightarrow \Lambda - \delta \Lambda$, the field is redefined according to $\varphi \rightarrow \varphi - \delta \Lambda /\Lambda \, \Psi$. This relationship between ERG equations and field redefinitions was first noticed by Wegner~\cite{Wegner_CS} and later explored by Latorre and Morris~\cite{TRM+JL}, in particular.

To summarize, equation~\eq{blocked} serves as our template for constructing ERG equations. Before diving into gauge theory, we will recall a particularly useful choice of $\Psi$ in scalar field theory, which we will subsequently generalize.

\subsection{Scalar Field Theory}

The flow equations we will deal with have the same basic structure as Polchinski's~\cite{Pol}. With this in mind, let us introduce a UV cutoff function $\cutoff(p^2/\Lambda^2)$, taken to die off rapidly for large argument and also exhibiting quasi-locality. The normalization is chosen such that $\cutoff(0) =1$. Next let us define an object which looks like a UV regularized propagator:
\be
	\ep_\Lambda(p^2) \equiv \frac{\cutoff(p^2/\Lambda^2)}{p^2}.
\ee
We will return to the interpretation of $\ep$ below, which can be subtle even in scalar field theory. It is now convenient to split up the action (which in this section we take as a functional of the scalar field, $\phi$) according to
\be
	\Stot_\Lambda[\phi] = \hS_\Lambda[\phi]  + \Sint_\Lambda[\phi]
\label{eq:split}
\ee
where, for the time being, we take 
\be
	\hS_\Lambda[\phi] = \hf \phi \cdot \ep^{-1}_\Lambda \cdot \phi = \MomInt{\D}{p} \phi(-p) \ep^{-1}_\Lambda(p^2) \phi(p)
.
\label{eq:seed-free}
\ee
Defining $\dot{\ep} \equiv -\Lambda d \ep_\Lambda /d \Lambda$ and 
\be
	\Sigma_\Lambda[\phi] \equiv \Stot_\Lambda[\phi] -2\hS_\Lambda[\phi],
\label{eq:Sigma}
\ee 
Polchinski's equation follows from the choice
\be
	\Psi(p) = \hf \dd_\Lambda(p^2) 
		\fder{\Sigma_\Lambda[\phi]}{\phi(-p)}	,
\label{eq:choice}
\ee
which yields the flow equation
\be
	-\flow \Stot_\Lambda[\phi] = \hf \classical{\Stot}{\dd_\Lambda}{\Sigma}{\phi} - 
	\hf \quantum{\dd_\Lambda}{\Sigma}{\phi}.
\label{eq:Pol}
\ee

We are now in a position to interpret $\ep$. First and foremost, note that $\dd$ is quasi-local and incorporates a UV cutoff function, giving a well defined flow equation. Indeed, from the point of view of the flow equation, it is $\dd$---referred to as an ERG kernel---and not $\ep$ which is the primitive object. \emph{After} solving the flow equation, we look at the two-point vertex of $\Stot$ and then interpret $\ep$ appropriately. For example if there are no contributions to the two-point vertex of $\Sint$ until $\order{p^4}$, then it makes sense to think of $\ep$ simply as the UV regularized propagator. However, it is quite permissible for $\Sint$ to contain a mass term, in which case we should change our interpretation appropriately. More severely, it is quite permissible for the solution to $\Sint$ to precisely cancel the $\order{p^2}$ part of the integrand in the `kinetic term' $\hS =  \hf \phi \cdot \ep^{-1}_\Lambda \cdot \phi$. Of course, the resulting theory will not be unitary after continuation to Minkowski space, but this is beside the point; rather, we wish to emphasise that it is \naive\ to simply state that $\ep_\Lambda$ is a UV regularized propagator. This issue will be sharpened in gauge theory where we will not fix the gauge and so cannot define the propagator!

Before moving on, let us note that it is simple to introduce source terms into the action, which is natural if one wants to compute correlation functions (of the fundamental field and/or composite operators). Indeed, we can achieve this simply by making the replacement $\Stot_\Lambda[\phi] \rightarrow \Stot_\Lambda[\phi,J]$, where $J$ stands for some set of sources. The corresponding operators to which these sources couple can be specified via a boundary condition for the bare action. It is easy to check that a source-dependent solution to~\eq{Pol} is
\be
	\Sint_\Lambda[\phi,J] = 
	-\MomInt{\D}{p}
	\biggl[
		J(p) \phi(p)
		+
		\hf 
		J(-p) \frac{1-\cutoff(p^2/\Lambda^2)}{p^2} J(p) 
	\biggr].
\ee
Part of the motivation for showing this solution is that it very nicely illustrates the various concepts of locality delineated earlier. First, recalling that $\cutoff(0)=1$, note that in the limit $\Lambda \rightarrow \infty$, the action becomes strictly local (and possesses the source term $J \cdot \phi$). For $0<\Lambda<\infty$, the action is quasi-local; this follows because, in the final term, $\cutoff(p^2/\Lambda^2) = 1 + \order{p^2/\Lambda^2}$. But in the limit $\Lambda \rightarrow 0$  the cutoff function disappears, yielding a non-local term. As alluded to above, this is hardly suprising. It has been shown in~\cite{Fundamentals} that $\lim_{\Lambda\rightarrow0} \Sint_\Lambda[0,J]$ generates the connected correlation functions of whatever is coupled in the UV. Here we are dealing with the Gaussian theory coupled to $J \cdot \phi$; and we simply recover the result that the two-point connected correlation function goes like $1/p^2$.

\subsection{Yang-Mills Theory}

\subsubsection{The Setup}

To construct ERGs for which gauge invariance is preserved along the flow, we require a gauge invariant cutoff.
A beautiful solution to this problem was found in~\cite{sunn} for $\SU(N)$ Yang-Mills. The idea is to embed the physical gauge theory into a spontaneously broken $\SU(N|N)$ gauge theory, regularized by covariant higher derivatives.  Were covariant higher derivative regularization applied solely to $\SU(N)$ Yang-Mills then, as has been known for a long time, a set of one-loop divergences slip through. However, $\SU(N|N)$ Yang-Mills has sufficiently improved UV properties to cure this.

The embedding is as follows. Denoting the physical gauge field by $A^1_\mu$ and the $\SU(N|N)$ super gauge field by $\mathcal{A}_\mu$, we can write
\be
	\mathcal{A}_\mu
	=
	\begin{pmatrix}
		A^1_\mu & B_\mu
	\\
		\bar{B}_\mu  & A^2_\mu
	\end{pmatrix}
	+ \mathcal{A}^0_\mu \mathbbm{1},
\ee
where $A^2_\mu$ carries an unphysical $\SU(N)$ theory (which turns out to come with a wrong-sign kinetic term), $B_\mu$ and $\bar{B}_\mu$ are wrong-statistics fermions and the final contribution is a central term.  Introducing a superscalar field, $\mathcal{C}$, with a vev such that the fermions acquire a mass, the $\SU(N|N)$ symmetry is broken down to $\SU(N) \times \SU(N) \times U(1)$. The symmetry breaking scale is chosen to be $\Lambda$, the same as the scale of the covariant higher derivatives. As shown in~\cite{sunn}, this scheme regularizes all Feynman graphs. Moreover, it was also established that the necessary decoupling properties are satisfied in order that a regularization of the physical $\SU(N)$ is achieved. 
To be precise, in the limit $\Lambda  \rightarrow \infty$ the fermions decouple. Furthermore, the lowest gauge invariant interaction between $A^1_\mu$ and $A^2_\mu$ is of the form $ \Int{x} \tr (F^1)^2 \tr (F^2)^2$, (where $F^{1,2}$ are the corresponding field strength tensors) which, on dimensional grounds, comes with a factor $1/\Lambda^{\D}$. Therefore, the unphysical $\SU(N)$ decouples from the physical one in the limit $\Lambda \rightarrow \infty$. Consequently, so long as we pose questions of only the physical sector (e.g.\ by computing the flow of the physical gauge coupling, or computing correlators involving only the physical field) then we are guaranteed that the effects of the unphysical fields merely amount to providing regularization. As for the central term, it turns out that $\mathcal{A}^0_\mu$ can and should be entirely removed from the action; the interested reader is referred to~\cite{mgierg1}. The regularization was extended to include quarks in~\cite{qcd}.

With the regularization scheme in place, we can now construct our flow equation. The idea is to use the template~\eq{Pol} but with several modifications appropriate to gauge theory. First, of all, we replace the field $\phi$ with $\varphi_i$, where $i$ runs over all fields in the broken phase of $\SU(N|N)$, embedded in a supermatrix. Thus, the component of $\varphi_i$ corresponding to $A^1$ stands for $\mathrm{diag} (A^1,0)$.
We must allow for a separate kernel, $\dd$, for each propagating field.
 To denote this, we label the kernels by the fields, so that e.g.\ $\dd^{A^1A^1}$ is the kernel appropriate to the physical gauge field. Next, we reinterpret $\hS$, which was previously given by~\eq{seed-free}. Returning to~\eq{blocked}, recall that we are free to make any choice of $\Psi$, so long as it leads to a well defined flow equation. Therefore, we are quite entitled to furnish $\hS$ with interactions, in order to render it gauge invariant. In actual fact, for various technical reasons discussed in~\cite{aprop,mgierg2}, it is necessary to go beyond the minimal choice and actually allow for a rather general $\hS$. With this done, we retain the definition of $\Sigma$ given in~\eq{Sigma}, noting that this object is now gauge invariant.

However, we are still not done. Notice that the flow equation involves the object
\be
	\fder{}{\varphi_i} \knl{\dd^{\varphi_i \varphi_j}} \fder{}{\varphi_j}
	=
	\Int{x}\volume{y}
	\fder{}{\varphi_i(x)} \dd^{\varphi_i \varphi_j}(x-y) \fder{}{\varphi_i(y)}
\ee
(with an implied sum over repeated indices)
which is not gauge invariant, due to the functional derivatives being at different points. This can be rectified by covariantizing the kernels, which essentially amounts to furnishing $\dd^{\varphi_i \varphi_j}(x-y)$ with vertices: just as the vertices of the Wilsonian effective action are related to each other by gauge invariance identities, so too are the vertices of the kernels. Denoting a covariantization via $\{\}$,
we can write (suppressing Euclidean indices)
\begin{multline}
	\fder{}{\varphi_i} \{\dd^{\varphi_i \varphi_j}\}\fder{}{\varphi_j} =
\\
	\sum_{n,m}
	\Int{x} \volume{y} \volume{x_1}\cdots \volume{x_n} \volume{y_1} \cdots \volume{y_m}
	\dd^{\varphi_{i_1}\cdots \varphi_{i_n}, \varphi_{j_1}\cdots \varphi_{i_m };\varphi_i \varphi_j}
	(x_1,\ldots,x_n,y_1,\ldots,y_m;x,y)
\\
	\str
	\biggl[
		\fder{}{\varphi_i(x)} \varphi_{i_1}(x_1)\cdots \varphi_{i_n}(x_n)
		\fder{}{\varphi_i(y)} \varphi_{j_1}(y_1)\cdots \varphi_{j_m}(y_m)
	\biggr],
\end{multline}
where $\str$ is the cyclic invariant for $\SU(N|N)$, consisting of the trace of the upper left $N\times N$ block of a supermatrix minus the trace of the lower right $N\times N$ block. The term with $n=m=0$ involves the underlying kernel and the higher
$\dd^{\varphi_{i_1}\cdots \varphi_{i_n}, \varphi_{j_1}\cdots \varphi_{i_m };\varphi_i \varphi_j}$ are the vertices which, beyond satisfying gauge invariance identities, can be taken to be general up to some restrictions described in~\cite{aprop}.

Before giving the flow equation, let us recall that it is convenient to rescale $\mathcal{A}_\mu \rightarrow \mathcal{A}_\mu/g$, so that the covariant derivative becomes
$\nabla_\mu = \partial_\mu - i \mathcal{A}_\mu$. Now, under gauge transformations, we have $\delta_\omega \mathcal{A}_\mu = [\nabla_\mu,\omega]$. Since we anticipate that we never fix the gauge, this is a manifest symmetry of the theory. Consequently, it is easy to check that $\mathcal{A}_\mu$ cannot renormalize, which would amount to sending $\mathcal{A}_\mu \rightarrow \mathcal{A}_\mu Z(\Lambda)$, as first noted in~\cite{ym1}. It turns out that this property is inherited by all fields in the broken phase~\cite{aprop,mgierg1}. There is, however, an apparent drawback: the rescaling $\mathcal{A}_\mu \rightarrow \mathcal{A}_\mu/g$ causes an extra term to appear on the left-hand side of the flow equation, after rewriting it in terms of $\flow$ taken at constant rescaled field, which is not manifestly gauge invariant~\cite{ym1}. The solution is to tune $\Psi$ to absorb this term. Finally, then, the flow equation takes the deceptively simple form
\be
	-\flow \Stot[\varphi] = 
	\hf \fder{\Stot}{\varphi_i} \{g^2 \dd^{\varphi_i \varphi_j}\}\fder{\Sigma}{\varphi_j} 
	-\hf \fder{}{\varphi_i} \{g^2 \dd^{\varphi_i \varphi_j}\}\fder{\Sigma}{\varphi_j} .
\label{eq:mgi-flow}
\ee
At no point in the construction of this equation has the gauge been fixed; the formalism is manifestly gauge invariant. Note that we can take this flow equation to define what we mean by the partition function in the absence of gauge fixing.

Let us now return to the issue of the interpretation of the integrated kernels. In the $A^1$ sector, we can choose $\dd^{A^1 A^1}$ to be essentially the same as for scalar field theory, corresponding to
\be
	\ep^{A^1 A^1}_{\mu \ \nu}(p^2) = 
	\delta_{\mu\nu} \frac{\cutoff(p^2/\Lambda^2)}{p^2}.
\ee
This looks like a regularized Feynman propagator, which is perhaps rather mysterious since we claim not to have fixed the gauge. However, consider focussing on the $A^1$ sector and pulling out the two-point term:
\be
	\Stot[\varphi] = 
	\frac{1}{g^2} A^1_\mu \cdot \bigl(D^{-1}\bigr)^{A^1 A^1}_{\mu\ \nu} \cdot A^1_\nu + \ldots
\ee
Substituting this into the flow equation, it is easy to check that
\be
	  \bigl(D^{-1}\bigr)^{A^1 A^1}_{\mu\ \alpha} (p)
	  \;
	  \ep^{A^1 A^1}_{\alpha \ \nu}(p^2) =
	 p^2 \delta_{\mu\nu} - p_\mu p_\nu.
\ee
Therefore, the integrated kernel is the inverse of the kinetic term in the transverse space. Thus despite appearances, the integrated kernel should not be interpreted as a propagator in the standard sense. Nevertheless, from a diagrammatic perspective~\cite{aprop,mgiuc} in particular it plays a rather similar role and so is often referred to as an effective propagator.

\section{Results to Date \& Outlook}

All quantitative work that has been done so far with the manifestly gauge invariant flow equation has been perturbative. Whilst the aim of constructing the formalism is to provide a nonperturbative tool, it is of course essential to test a local, continuum approach that makes so grand a claim as to be manifestly gauge invariant. Even before the $\SU(N|N)$ regularization scheme was fully understood, it was possible to compute the one-loop $\beta$-function in pure $\SU(N)$ Yang-Mills~\cite{ym2}. Subsequent to the details of the regularization being fleshed out, this computation was redone and greatly refined~\cite{aprop}. The main improvement was to recognize that much of the calculation could be done without ever specifying the details of either the $\hS$ or the covariantization of the kernels. These techniques were further refined~\cite{mgierg1}, paving the way for the successful computation of the two-loop result in~\cite{mgierg2}, thereby providing a highly non-trivial test of the formalism.

Nevertheless, despite the enhanced understanding of the formalism, the two-loop calculation was something of a nightmare, involving as it did the generation of thousands of terms, most of which cancelled identically to leave a small set from which the final answer was extracted. Subsequently, it was realised that many of these cancellations can be done in parallel~\cite{Primer}; following on from this, a diagrammatic expression was derived, to all loops, for the surviving terms~\cite{mgiuc}. This represents a colossal improvement over the original two-loop calculation and, indeed, reduces the complexity to something similar to a more standard calculation. Building on this, it was shown that the expectation value of a long, thin, rectangular Wilson loop can be easily computed without gauge fixing~\cite{evalues}. When the formalism was extended to QCD~\cite{qcd}, the one-loop $\beta$-function was computed with the minimum of fuss.

Up to this point, the simplifications uncovered were rooted in perturbation theory. A step towards going beyond this was given in~\cite{univ}, representing the last piece of work, to date, using the formalism. Recently, however, there have been substantial advances in the understanding of the structure of flow equations~\cite{Fundamentals}. In the context of scalar field theory, this has allowed several results to be derived without any truncation (or diagrammatic expansion), including a demonstration that the spectrum of the anomalous dimension at critical fixed-points is quantized~\cite{Fundamentals} and a proof of an extension of Pohlmeyer's theorem~\cite{OJR-Pohl}. It is clear that these insights into the flow equation can be used to place~\cite{univ} on a proper footing, that no longer makes any reference to diagrammatics.

Coming from a somewhat different direction, a compelling qualitative picture for a confinement mechanism has been presented in~\cite{conf}. As part of this, it is observed that in any manifestly gauge invariant formulation of Yang-Mills, the gluon cannot have a mass gap. Intuitively, then, one might expect that, \emph{in this manifestly gauge invariant picture}, confinement is driven by $g$ diverging in the IR. This suggests that it might be profitable to perform an expansion in $1/g^2$ but where it should be emphasised that this $g$ is the renormalized and not the bare coupling. As discussed in~\cite{conf}, the scheme has attractive features. However, there is a severe problem: notice the explicit factors of $g^2$ appearing in the flow equation~\eq{mgi-flow}; whilst they guarantee closure of perturbation theory, they have the opposite effect in a strong coupling expansion. Therefore, to use the strong coupling expansion, either some resummation or further approximation must be performed.

There has been no investigation into either of these choices, partly due to the daunting complexity of the flow equation once one looks under the bonnet. However, as intimated above, the understanding of the flow equation is much better than when these issues were first considered. This is, therefore, the right time to return to the manifestly gauge invariant ERG, aiming to combine the physical insights made some time ago with the technical advances made more recently.


\end{document}